\documentstyle[12pt]{article}
\begin{document}
\begin{center}
\centerline{\bf ADELIC PATH INTEGRALS FOR QUADRATIC LAGRANGIANS}

\bigskip
\bigskip
\centerline{ GORAN S. DJORDJEVI\'C}
{\it Department of Physics, Faculty of Sciences, University of Ni\v s,
P.O.Box 224, 18001 Ni\v s, Yugoslavia}
\centerline{\it and}
{\it Sektion Physik, Universit\"at M\"unchen, Theresienstr. 37,
D-80333 M\"unchen, Germany}

\bigskip
\centerline{   BRANKO  DRAGOVICH }
{\it Steklov Mathematical Institute, Gubkin St. 8, GSP-1, 117966,
Moscow, Russia}
\centerline{\it and}
{\it Institute of Physics, P.O.Box 57, 11001 Belgrade, Yugoslavia }

\bigskip
\centerline{ LJUBI\v SA  NE\v SI\'C}
{\it Department of Physics, Faculty of Sciences, University of Ni\v s,
P.O.Box 224, 18001 Ni\v s, Yugoslavia}
\date{}
\end{center}

\begin{abstract} %Here an Abstract.
Feynman's path integral in adelic quantum mechanics is considered.
The propagator ${\cal K}(x^{''},t^{''};x',t')$ for one-dimensional
adelic systems with quadratic Lagrangians is analytically
evaluated. Obtained exact general formula has the  form which is
invariant under interchange of the number fields $\bf R$ and ${\bf Q}_p$.
\end{abstract}

\noindent
{\bf 1. Introduction}
\smallskip

\noindent
For simplicity, we consider one-dimensional systems, but many issues
can be easily generalized to higher dimensions.

It is well known that dynamical evolution of any
quantum-mechanical system, described by a wave function $\Psi (x,t)$,
is given by
$$
   \Psi (x'',t'') = \int {\cal K}(x'',t'';x',t')\Psi(x',t')dx' , \eqno(1.1)
$$
where ${\cal K}(x'',t'';x',t')$  is the kernel of the corresponding unitary
integral operator acting as follows:
$$
   \Psi (t'') =  U(t'',t')\Psi (t') .              \eqno(1.2)
$$

$ {\cal K}(x'',t'';x',t')$ is also called Green's function, or the
quantum-mechanical propagator, and the probability amplitude to go a
particle from a space-time point $ (x',t')$ to the other point
$(x'',t'')$. Starting from (1.1) one can easily derive the following
three general properties:
$$
  \int{\cal K}(x'',t'';x,t) {\cal K}(x,t;x',t') dx =
{\cal K}(x'',t'';x',t') ,      \eqno(1.3)
$$
$$
   \int \bar{{\cal K}}(x'',t'';x',t') {\cal K}(y,t'';x',t') dx' =
\delta (x''-y) ,       \eqno(1.4)
$$
$$
   {\cal K}(x'',t'';x',t'') = \lim_{t'\to t''} {\cal K}(x'',t'';x',t') =
\delta (x''-x') .    \eqno(1.5)
$$
Since all information on quantum dynamics can be deduced from
the propagator ${\cal K}(x'',t'';x',t')$ it can be regarded as
the basic ingredient of quantum theory. In Feynman's formulation
\cite{feynm1} of quantum mechanics,
${\cal K}(x'',t'';x',t')$ was postulated to be the path integral
$$
   {\cal K}(x'',t'';x',t') = \int_{x',t'}^{x'',t''}
   \exp{\left(\frac{2\pi i}{h} \int_{t'}^{t''}
   L(\dot{q},q,t)dt\right)} {\cal D}q ,  \eqno(1.6)
$$
where $x'' = q(t'')$ and $x' = q(t')$, and $h$ is the Planck constant.

In its original form, the path integral (1.6) is the limit of
the corresponding multiple integral of $n$ variables
$q_i = q(t_i), \ \   (i=1,2,...,n),$ when $n\to\infty$.
For all its history, the path integral
has been a subject of permanent interest in theoretical
and mathematical physics. At present days (see, e.g. \cite{firen1}),
it is one of the most profound and promising approaches to
foundations of quantum theory (in particular, quantum field theory
and superstring theory).
Feynman's path integral is a natural instrument
in formulation and investigation of $p$-adic \cite{volov1}
and adelic \cite{brank1} quantum mechanics.

Adelic quantum mechanics, we are interested in, contains
complex-valued functions of real and all $p$-adic arguments.
There is not the
corresponding Schr$\ddot{o}$dinger equation, but Feynman's
path integral method is quite appropriate. Feynman's path integral for $p$-adic
propagator ${\cal K}_p(x'',t'';x',t')$ \cite{volov1}, where ${\cal K}_p$ is
complex-valued and $x'',x',t'',t'$
are $p$-adic variables, is a direct $p$-adic generalization of (1.6),
i.e.
$$
   {\cal K}_p (x'',t'';x',t') = \int_{x',t'}^{x'',t''}
   \chi_p \left( -\frac{1}{h}
   \int_{t'}^{t''} L(\dot{q},q,t)dt\right) {\cal D}q ,    \eqno(1.7)
$$
where $ \chi_p(a)$ is $p$-adic additive character.
The Planck constant $h$ in (1.6) and (1.7) is the same rational number.
We consider integral $ \int_{t'}^{t''}L(\dot{q},q,t)dt$  as
the difference of antiderivative (without pseudoconstants) of
$L(\dot{q},q,t) $ in final $(t'')$ and initial $(t')$ times.
${\cal D}q = \prod_{i=1}^n dq(t_i)$, where $dq(t_i)$ is the $p$-adic
Haar measure. Thus, $p$-adic
path integral is the limit of the multiple Haar integral when
$n\to\infty$. To calculate (1.7) in this way  one has to introduce
some order on $t\in Q_p$, and it is successfully done in Ref.
\cite{zelen1}. On previous investigations of $p$-adic path
integral one can see  \cite{brank2,brank3}. Our main task here  is
a derivation of the exact result for $p$-adic (1.7) and the
corresponding adelic Feynman path integral for the general
case of Lagrangians
$L(\dot{q},q,t)$, which are quadratic polynomials in $\dot{q}$
and $q$, without making time discretization. In fact, we will use
the general requirements (in particular, (1.3) and (1.4))
which any  (ordinary, $p$-adic
and adelic) path integral must  satisfy. In some parts of evaluation,
we exploit methods inspired by Ref. \cite{grosj1}.

Adelic path integral
may be regarded as an infinite product of the ordinary one and
$p$-adic path integrals for all primes $p$. Formal definition
of adelic functional integral, with some of its basic properties,
will be presented in Section 5.

Since 1987, there have been many publications (for a review,
see, e.g. \cite{volov2,freun1,khren1} ) on possible applications
of $p$-adic numbers and adeles in  modern theoretical and
mathematical physics. The first successful employment of
$p$-adic numbers was in string theory. In Volovich's article
\cite{volov3}, a hypothesis on the existence of nonarchimedean
geometry at the Planck scale was proposed and $p$-adic string theory
was initiated. In particular, it was proposed generalization of the
standard Veneziano and Virasoro-Shapiro amplitudes with
complex-valued multiplicative characters over various number
fields, and $p$-adic valued Veneziano amplitude was constructed
by means of interpolation. Using $p$-adic Veneziano amplitude
as the Gel'fand-Graev \cite{gelfa1} beta function, Freund
and Witten obtained \cite{freun2} an attractive adelic
formula, which states that the product of the standard crossing
symmetric Veneziano amplitude and all its $p$-adic counterparts
equals unit. Such approach gives a possibility to consider
some ordinary string amplitudes as an infinite product of
their inverse $p$-adic analogues. Many $p$-adic aspects of
string theory and nonarchimedean geometry of the space-time at the
Planck scale have been of the significant interest during the
last fifteen years.

For a systematic investigation of $p$-adic quantum dynamics, two kinds
of $p$-adic quantum mechanics are formulated: with complex-valued
and $p$-adic valued wave functions of $p$-adic variables
(for a review, see \cite{volov1,volov2} and \cite{khren1},
respectively). This paper is related to the first kind, which
can be presented as a triple
$$
    (L_2 ({\bf Q}_p), W, U(t)),                      \eqno(1.8)
$$
where $L_2({\bf Q}_p)$ is the Hilbert space on ${\bf Q}_p$,
$W$ denotes the Weyl quantization procedure, and
$U(t)$ is the unitary representation of an evolution operator
on $L_2({\bf Q}_p)$. In our approach, $U(t)$ is naturally realized
by the Feynman path integral method. In order to connect
$p$-adic with standard quantum mechanics, adelic quantum mechanics
is formulated \cite{brank1}. Within adelic quantum mechanics
a few basic physical systems, including some minisuperspace
cosmological models, have been successfully considered.
As a result of $p$-adic effects in the adelic approach,
a space-time discretness at the Planck scale is obtained.
Adelic path integral plays a central role and provides
an extension of possible quantum trajectories over real space
to all $p$-adic spaces. There has been also application of
$p$-adic numbers in the investigation of spin glasses, Brownian motion,
stochastic processes, information systems, hierarchy structures
and some other problems related to very complex dynamical
systems.

Before to proceed with path integral, let us give a short review
of some basic properties of $p$-adic numbers, adeles, and analyses over them,
which provide a minimum of necessary mathematical background.
It will also contain mathematical concepts to present a
motivation for employment of $p$-adic numbers and adeles in physics.

\bigskip

\noindent
{\bf 2. $p$-Adic and Adelic Analyses, and Motivations for Their Use
in Physics}
\smallskip

\noindent
There are physical and mathematical reasons to start with the field
of rational numbers ${\bf Q}$. From physical point of view, numerical
results of all experiments and observations are some rational numbers,
{\it i.e.} they belong to ${\bf Q}$. From algebraic
point of view, ${\bf Q}$ is the simplest number field of characteristic
$0$. Recall that any $0\neq x\in {\bf Q}$ can be presented as infinite
expansions into the two essentially different forms:
$$
 x = \sum_{k=n}^{-\infty} a_k 10^k, \ \  a_k = 0,1,\cdots ,9,
\ \ a_n \neq 0,   \eqno(2.1)
$$
which is the ordinary one to the base $10$, and the other one to the base
$p$ ($p$ is any prime number)
$$
  x =  \sum_{k=m}^{+\infty} b_k p^k, \ \  b_k = 0,1,\cdots ,p-1,
\ \  b_m \neq 0,  \eqno(2.2)
$$
where $n$ and   $m$ are some integers which depend on $x$.
The above representations (2.1) and  (2.2) exhibit the usual repetition
of digits, however the expansions are in the mutually opposite directions.
The series (2.1) and  (2.2) are convergent with respect to the metrics
induced by the usual absolute value $| \cdot |_\infty$ and
$p$-adic absolute value (or $p$-adic norm) $| \cdot |_p$, respectively.
Note that these valuations exhaust all possible inequivalent non-trivial
norms on ${\bf Q}$.
Allowing all possible combinations for digits, one obtains
standard representation of real and $p$-adic numbers in the form
(2.1) and (2.2), respectively. Thus, the field of real numbers
${\bf R}$ and the fields of $p$-adic numbers ${\bf Q}_p$
exhaust all number fields
which may be obtained by completion of ${\bf Q}$, and which contain
${\bf Q}$ as a dense subfield. Since $p$-adic norm of any term in (2.2)
is $| b_k p^k |_p = p^{-k}$ if $b_k \neq 0$, geometry of $p$-adic numbers is
the nonarchimedean one, {\it i.e.} strong triangle inequality
$| x+y |_p \leq max
(| x |_p, | y |_p) $ holds and $| x |_p =p^{-n}$.
${\bf R}$ and ${\bf Q}_p$ have
many distinct algebraic and geometric properties.

There is no natural ordering on ${\bf Q}_p$. However one can introduce
a linear order on ${\bf Q}_p$ in the following way: $x < y$ if
$| x |_p < | y|_p$, or if $| x|_p = | y|_p$
then there exists such index $r \geq 0$ that digits satisfy
$x_m = y_m, x_{m+1} = y_{m+1}, \cdots, x_{m+r-1} = y_{m+r-1},
x_{m+r} < y_{m+r}$. Here, $x_k$ and $y_k$ are digits related to $x$ and $y$
in expansion (2.2). This ordering is very useful in time discretization
and calculation of $p$-adic functional integral as a limit of
the  $n$-multiple Haar integral when $n\to \infty$.

There are mainly two kinds of analysis on ${\bf Q}_p$ which are of interest
for physics, and they are based on two different mappings: ${\bf Q}_p
\to {\bf Q}_p$ and ${\bf Q}_p\to {\bf C}$, where ${\bf C}$ is the
field of ordinary complex numbers. We use both of these analyses, in classical
and quantum $p$-adic models, respectively.

Elementary $p$-adic valued functions and their derivatives are defined
by the same series as in the real case, but their regions of convergence
are determined by means of $p$-adic norm. As a definite $p$-adic valued
integral of an analytic function $f(x) = f_0 + f_1 x + f_2 x^2 + \cdots$
we take difference of the corresponding antiderivative in end points, i.e.
$$
 \int_a^b f(x) = \sum_{n=0}^\infty \frac{f_n}{n+1} \left( b^{n+1} -
 a^{n+1}   \right).
$$

Usual complex-valued functions of $p$-adic variable are: ({\it i})
an additive character $\chi_p (x) = \exp{2\pi i \{x \}_p}$, where
$\{x \}_p$ is the fractional part of $x\in {\bf Q}_p$, ({\it ii})
a multiplicative character $\pi_s (x) = |x|_p^s$, where $s\in {\bf C}$,
and ({\it iii}) locally constant functions with compact support, like
$\Omega (|x|_p)$, where
$$
\Omega (|x|_p) = \left\{  \begin{array}{ll}
                 1,   &   |x|_p \leq 1,  \\
                 0,   &   |x|_p > 1.
                 \end{array}    \right.                     \eqno(2.3)
$$
There is well defined Haar measure and integration, and we use
the Gauss integral
$$
  \int_{{\bf Q}_p} \chi_p (\alpha x^2 + \beta x) dx
  =\lambda_p (\alpha) |2\alpha|_p^{-\frac{1}{2}}
  \chi_p \left( -\frac{\beta^2}{4\alpha} \right), \ \ \alpha\neq 0.
                                                   \eqno(2.4)
$$
The arithmetic function $\lambda_p (x)$ in (2.4) is a map
$\lambda_p: {\bf Q}_p^\ast\to {\bf C}$ defined as follows
\cite{volov2}:
$$
 \lambda_p (x) = \left\{   \begin{array}{ll}
                 1,    &  m=2j,  \ \ \ \ \ \ \  p\neq 2,  \\
                 \left(\frac{x_m}{p} \right),    &  m=2j+1,
 \ \  p\equiv 1(\mbox{mod}\ 4), \\
                 i\left(\frac{x_m}{p} \right),    &  m=2j+1,
  \ \  p\equiv 3(\mbox{mod}\ 4),
                 \end{array}   \right.            \eqno(2.5)
$$
$$
  \lambda_2 (x) = \left\{   \begin{array}{ll}
                  \frac{1}{\sqrt 2} [1 + (-1)^{x_{m+1}} i],
   &  m= 2j,  \\
                 \frac{1}{\sqrt 2} (-1)^{x_{m+1}+ x_{m+2}}
                 [1 + (-1)^{x_{m+1}} i],      &   m=2j+1,
                 \end{array}
                 \right.        \eqno(2.6)
$$
where $x$ is presented in the form (2.2),  $j\in {\bf Z}$,
 $\left(\frac{x_m}{p} \right)$ is the  Legendre symbol and
${\bf Q}_p = {\bf Q}_p\setminus \{0\}$.
It is often enough to use properties:
$$
  \lambda_p (a^2 x)= \lambda_p (x), \ \  \lambda_p (x) \lambda_p (-x) =1,
\ \  \lambda_p (x)\lambda_p (y)  = \lambda_p (x+y)  \lambda_p (x^{-1}+y^{-1}),
$$
$$
|\lambda_p (x)|_\infty = 1,   \ \ a\neq 0.      \eqno(2.7)
$$
For a more complete information on $p$-adic numbers and
nonarchimedean analysis one can see \cite{volov2,gelfa1,schik1}.

It is worth noting that the real analogue of the Gauss integral
(2.4)  has the same form, {\it i.e.}
$$
  \int_{Q_\infty} \chi_\infty (\alpha x^2 + \beta x) dx
  =\lambda_\infty (\alpha) |2\alpha|_\infty^{-\frac{1}{2}}
  \chi_\infty \left( -\frac{\beta^2}{4\alpha} \right), \ \ \alpha\neq 0,
                                                   \eqno(2.8)
$$
where ${\bf Q_\infty}\equiv {\bf R}$ and $\chi_\infty (x) =
\exp{(-2\pi i x)}$ is additive character in the real case.
Function $\lambda_\infty (x)$ is defined as
$$
  \lambda_\infty (x) = \sqrt{\frac{\mbox{sign}\ x}{i}}
=\exp{\left( -i\frac{\pi}{4}\right)}\sqrt{\mbox{sign}\ x},
\ \ x\in {\bf R}^\ast  \eqno(2.9)
$$
and exhibits the same properties (2.7).

Real and $p$-adic numbers are unified in the form of adeles.
An adele $x$ \cite{gelfa1} is an infinite sequence
$$
  x= (x_\infty, x_2, \cdots, x_p, \cdots),             \eqno(2.10)
$$
where $x_\infty \in {\bf R}$ and $x_p \in {\bf Q}_p$ with
the restriction that for all but a finite set $\bf S$ of primes
$p$ one has  $x_p \in {\bf Z}_p$, where ${\bf Z}_p = \{ a\in {\bf Q}_p:
|a|_p \leq 1  \}$ is the ring of $p$-adic integers.
Componentwise addition and multiplication are natural operations
on the ring of adeles ${\cal A}$, which can be regarded as
$$
 {\cal A} = \bigcup_{{\bf S}} {\cal A} ({\bf S}),
 \ \  {\cal A}({\bf S}) = {\bf R}\times \prod_{p\in {\bf S}} {\bf Q}_p
 \times \prod_{p\not\in {\bf S}} {\bf Z}_p.         \eqno(2.11)
$$
${\cal A}$ is also locally compact topological space.

There are also two kinds of analysis over topological ring of adeles
${\cal A}$, which are generalizations of the corresponding analyses
over $\bf R$ and  ${\bf Q}_p$. The first one is related to mapping
${\cal A}\to {\cal A}$ and the other one to ${\cal A}\to {\bf C}$.
In complex-valued adelic analysis it is worth mentioning  an
additive character
$$
 \chi (x) = \chi_\infty (x_\infty) \prod_p \chi_p (x_p),   \eqno(2.12)
$$
a multiplicative character
$$
  |x|^s = |x_\infty|_\infty^s \prod_p |x_p|_p^s, \ \ s\in {\bf C},
                                                    \eqno(2.13)
$$
and elementary functions of the form
$$
 \phi (x) = \phi_\infty (x_\infty) \prod_{p\in {\bf S}} \phi_p (x_p)
 \prod_{p\not\in {\bf S}} \Omega (|x_p|_p),           \eqno(2.14)
$$
where $\phi_\infty (x_\infty)$ is an infinitely differentiable
function on ${\bf R}$ such that $|x_\infty |_\infty^n
\phi_\infty (x_\infty)
\to 0$ as $|x_\infty|_\infty \to \infty$ for any $n\in
\{0,1,2,\cdots  \}$, and $\phi_p (x_p)$ is a locally constant function
with compact support. All finite linear combinations of elementary
functions (2.14) make up the set $S({\cal A})$ of the Schwartz-Bruhat
adelic functions. The Fourier transform of $\phi (x)\in S({\cal A})$,
which maps $S(\cal A)$ onto ${\cal A}$, is
$$
 \tilde{\phi}(y) = \int_{\cal A} \phi (x)\chi (xy)dx,      \eqno(2.15)
$$
where $\chi (xy)$ is defined by (2.12) and $dx = dx_\infty dx_2 dx_3
\cdots$ is the Haar measure on ${\cal A}$.

One can define the Hilbert space on ${\cal A}$, which we will denote
by $L_2({\cal A})$. It contains infinitely many complex-valued
functions of adelic argument (for example, $\Psi_1(x), \Psi_2(x), \cdots$)
with scalar product
$$
 (\Psi_1,\Psi_2) = \int_{\cal A} \bar{\Psi}_1(x) \Psi_2(x) dx
$$
and norm
$$
  ||\Psi|| = (\Psi,\Psi)^{\frac{1}{2}} < \infty,
$$
where $dx$ is the Haar measure on ${\cal A}$. A basis of $L_2({\cal A})$
may be given by the set of orthonormal eigefunctions in spectral problem
of the evolution operator $U (t)$, where $t\in {\cal A}$. Such
eigenfunctions have the form
$$
 \psi_{{\bf S},\alpha} (x,t) = \psi_n^{(\infty)}(x_\infty,t_\infty)
 \prod_{p\in {\bf S}} \psi_{\alpha_p}^{(p)} (x_p,t_p)
 \prod_{p\not\in {\bf S}} \Omega (|x_p|_p),                    \eqno(2.16)
$$
where $\psi_n^{(\infty)}$ and $\psi_{\alpha_p}^{(p)}$ are eigenfunctions
in ordinary and $p$-adic cases, respectively. $\Omega (|x_p|_p)$ is
defined by (2.3) and presents a vector invariant under transformation of
$U_p(t_p)$ evolution operator.
Adelic quantum mechanics \cite{brank1} may be regarded as a triple
$$
    (L_2 ({\cal A}), W(z), U(t)),                           \eqno(2.17)
$$
where $W(z)$  and $U(t)$ are unitary representations of the
Heisenberg-Weyl group and evolution operator on $L_2 ({\cal A})$,
respectively.

Since ${\bf Q}$ is dense not only in ${\bf R}$ but also in
${\bf Q}_p$ there is a sense to ask a question: Why real (and complex) numbers
are so good in description of usual physical phenomena, and, is there
any aspect of physical reality which cannot be described without $p$-adic
numbers? As adeles contain real and $p$-adic numbers, and
consequently archimedean and nonarchimedean geometries,
it has been natural to formulate  adelic quantum mechanics,
which provides a suitable framework for a systematic investigation
of the above question. Uncertainty in space measurements
$$
 \Delta x \Delta y \geq \ell_0^2 = \frac{\hbar G}{c^3}
 \sim 10^{-66} cm^2,                                            \eqno(2.18)
$$
which comes from ordinary quantum gravity considerations
based on real numbers, restricts application of ${\bf R}$
and gives rise for employment of ${\bf Q}_p$ at the Planck scale.
It is also reasonable to expect that fundamental physical laws are
invariant under interchange of ${\bf R}$ and ${\bf Q}_p$
\cite{volov4}.

\bigskip
\noindent
{\bf 3. Quadratic Lagrangians and Related Classical Actions}
\smallskip

\noindent
A general quadratic Lagrangian can be written as follows:
$$
   L(\dot{q},q,t) = \frac{1}{2}\frac{\partial^2 L_0}{\partial\dot{q}^2}\dot{q}^2
   +   \frac{\partial^2 L_0}{\partial\dot{q}\partial q}\dot{q}q
   +   \frac{1}{2}\frac{\partial^2 L_0}{\partial q^2} q^2
   +   \frac{\partial L_0}{\partial\dot{q}} \dot{q}
   +   \frac{\partial L_0}{\partial q} q  +  L_0,                \eqno(3.1)
$$
where index $0$ denotes that the Taylor expansion of $L(\dot{q},q,t)$
is around $\dot{q}=q=0$. In fact, we want to consider the corresponding
adelic Lagrangian, {\it i.e.} an adelic collection of Lagrangians of
the same form (3.1) which differ only by their valuations
$v = \infty, 2, 3,\cdots$. In this Section we  present some results
valid simultaneously for real as well as for $p$-adic classical mechanics.
Coefficients in the expansion (3.1) will be regarded as analytic functions of
the time $t$, where $\frac{\partial^2 L_0}{\partial\dot{q}^2}|_{t=0}\neq 0$,
and which power series have the same rational coefficients in the real
and all $p$-adic cases.

The Euler-Lagrange equation of motion is
$$
 \frac{\partial^2 L_0}{\partial\dot{q}^2} \ddot{q} +
 \frac{d}{dt} \left( \frac{\partial^2 L_0}{\partial\dot{q}^2} \right) \dot{q} +
 \left[ \frac{d}{dt} \left( \frac{\partial^2 L_0}{\partial\dot{q}\partial q}  \right) -
 \frac{\partial^2 L_0}{\partial q^2}  \right] q = \frac{\partial L_0}{\partial q}
 - \frac{d}{dt}\left( \frac{\partial L_0}{\partial\dot{q}}  \right) .
 \eqno(3.2)
$$
A general solution of (3.2) describes classical trajectory
$$
q = x(t) = C_1 f_1(t) + C_2 f_2(t) + \xi (t) ,   \eqno(3.3)
$$
where $f_1(t)$ and $f_2(t)$ are two linearly independent solutions
of the corresponding homogeneous equation, and $\xi (t)$ is a particular
solution of the complete equation (3.2). Imposing the boundary conditions
$x'=x(t')$ and    $x''=x(t'')$, constants of integration $C_1$ and  $C_2$
become:
$$
   C_1 \equiv C_1 (t'',t') = \frac{(x'-\xi')f''_2 - (x''-\xi'')f'_2}
{f''_2 f'_1 - f''_1 f'_2},     \eqno(3.4)
$$
$$
   C_2 \equiv C_2 (t'',t')  =  \frac{(x''-\xi'')f'_1 - (x'-\xi')f''_1}
{f''_2f'_1 - f''_1f'_2},   \eqno(3.5)
$$
where $f'_1 = f_1(t'), f''_2 = f_2(t'')$ and $\xi' =\xi(t'),
\xi'' = \xi(t'')$. Note that not all boundary values
$x''$ and $x'$ are possible. Namely, if $f''_2 f'_1 -
f''_1 f'_2 = 0$ then must be $x''-\xi'' = x'-\xi'$. Such
situation is typical for the case of periodic solutions
$f_1(t)$ and  $f_2(t)$.

Using the equation of motion (3.2), the Lagrangian (3.1) can be rewritten
as
$$
  L(\dot{x},x,t) = \frac{1}{2}\frac{d}{dt}
  \left[ \frac{\partial^2L_0}{\partial\dot{x}^2} \dot{x}x +
  \frac{\partial^2L_0}{\partial\dot{x}\partial x} x^2 +
  \frac{\partial L_0}{\partial\dot{x}} x   \right] +
  \frac{1}{2} \left( \frac{\partial L_0}{\partial\dot{x}} \dot{x} +
  \frac{\partial L_0}{\partial x} x  \right) + L_0
                                                              \eqno(3.6)
$$
where $x(t)$ is related to the classical trajectory (3.3).
Since $C_1(t'',t')$ and   $C_2(t'',t')$ are linear in $x''$  and $x'$,
the corresponding classical action
$\bar{S}(x'',t'';x',t') = \int_{t'}^{t''} L(\dot{x},x,t)dt$
is quadratic in $x''$ and
$x'$. Performing explicit integration for the first term of (3.6),
and taking into account (3.4) and (3.5), we get
$$
 \bar{S}(x'',t'';x',t') = \frac{1}{2}\frac{\partial^2\bar{S}_0}
 {\partial x''^2} x''^2 + \frac{\partial^2 \bar{S}_0}{\partial x''
  \partial x'} x'' x' + \frac{1}{2}\frac{\partial^2 \bar{S}_0}{\partial
  x'^2} x'^2  + \frac{\partial\bar{S}_0}{\partial x''} x''   +
  \frac{\partial\bar{S}_0}{\partial x'} x' + \bar{S}_0,
                                                         \eqno(3.7)
$$
where:
$$
  \frac{\partial^2\bar{S}_0}{\partial x''^2} =
  \frac{\partial^2 L_0''}{\partial\dot{x}^2} \frac{\dot{f}_2'' f_1'
  - \dot{f}_1'' f_2'}{\Delta (t'',t')} + \frac{\partial^2 L_0''}
  {\partial\dot{x}\partial x},                        \eqno(3.8)
$$
$$
  \frac{\partial^2 \bar{S}_0}{\partial x''\partial x'} =
 \frac{1}{2} \frac{\partial^2 L_0''}{\partial\dot{x}^2} \frac
 {f_2''\dot{f}_1'' - \dot{f}_2''f_1''}{\Delta(t'',t')} +
 \frac{1}{2} \frac{\partial^2 L_0'}{\partial\dot{x}^2} \frac
 {f_2'\dot{f}_1' - \dot{f}_2'f_1'}{\Delta(t'',t')},     \eqno(3.9)
$$
$$
  \frac{\partial^2\bar{S}_0}{\partial x'^2} =
  \frac{\partial^2 L_0'}{\partial\dot{x}^2} \frac{\dot{f}_2' f_1''
  - \dot{f}_1' f_2''}{\Delta (t'',t')} - \frac{\partial^2 L_0'}
  {\partial\dot{x}\partial x},                         \eqno(3.10)
$$
$$
  \frac{\partial\bar{S}_0}{\partial x''} =
\frac{1}{2} \left[
  \frac{\partial^2 L_0''}{\partial\dot{x}^2}
  \frac{\xi'' (f_2'\dot{f}_1'' - \dot{f}_2''f_1') + \xi'
 (\dot{f}_2''f_1'' - \dot{f}_1''f_2'')}{\Delta(t'',t')} +\dot{\xi}''
+\frac{\partial L_0''}{\partial x}\right]
$$
$$
+ \frac{1}{2}\left[\frac{f_1'}{\Delta(t'',t')}
 \int_{t'}^{t''}(\frac{\partial L_0}{\partial\dot{x}}\dot{f}_2 +
 \frac{\partial L_0}{\partial x} f_2) dt
- \frac{f_2'}{\Delta(t'',t')} \int_{t'}^{t''}(\frac{\partial L_0}
{\partial\dot{x}}\dot{f}_1 + \frac{\partial L_0}{\partial x} f_1) dt
  \right],                                              \eqno(3.11)
$$
$$
  \frac{\partial\bar{S}_0}{\partial x'} =
\frac{1}{2} \left[
  \frac{\partial^2 L_0'}{\partial\dot{x}^2}
  \frac{\xi' (f_2''\dot{f}_1' - \dot{f}_2'f_1'') + \xi''
 (\dot{f}_2'f_1' - \dot{f}_1'f_2')}{\Delta(t'',t')} -\dot{\xi}'
- \frac{\partial L_0'}{\partial x}\right]
$$
$$
+ \frac{1}{2}\left[\frac{f_2''}{\Delta(t'',t')}
 \int_{t'}^{t''}(\frac{\partial L_0}{\partial\dot{x}}\dot{f}_1 +
 \frac{\partial L_0}{\partial x} f_1) dt
- \frac{f_1''}{\Delta(t'',t')} \int_{t'}^{t''}(\frac{\partial L_0}
{\partial\dot{x}}\dot{f}_2 + \frac{\partial L_0}{\partial x} f_2) dt
  \right],                                              \eqno(3.12)
$$
$$
  \bar{S}_0 = \frac{1}{2}
\frac{\xi''f_2' - \xi' f_2''} {\Delta(t'',t')} \int_{t'}^{t''}
(\frac{\partial L_0}{\partial\dot{x}}\dot{f}_1 +
\frac{\partial L_0}{\partial x} f_1) dt
$$
$$
+\frac{1}{2}\frac{\xi'f_1'' - \xi'' f_1'} {\Delta(t'',t')}
\int_{t'}^{t''}(\frac{\partial L_0}{\partial\dot{x}}\dot{f}_2 +
\frac{\partial L_0}{\partial x} f_2) dt
$$
$$
+\frac{1}{2}\int_{t'}^{t''}(\frac{\partial L_0}{\partial\dot{x}}\dot{\xi} +
\frac{\partial L_0}{\partial x} \xi ) dt   +
\int_{t'}^{t''} L_0 dt.
                                                       \eqno(3.13)
$$
The expression (3.7) is presented in the form of the Taylor expansion
of the classical action around $x'' = x' =0$, where coefficient functions
depend on $t''$ and $t'$.
In the above expressions we denoted $L_0'' = L_0 (t'')$,
$L_0' = L_0 (t') $, and
$$
 \Delta (t'',t') = f_2''f_1' - f_2'f_1''.            \eqno(3.14)
$$
Thus, to Lagrangian $L(\dot{q},q,t)$, which is at most quadratic in
$\dot{q}$ and   $q$, corresponds classical action $\bar{S}(x'',t'';x',t')$
at most quadratic in $x''$  and  $x'$.

It is worth noting that solutions $f_2 (t)$ and $\xi (t)$ can be
expressed by means of solution $f_1 (x)$ in the following way:
$$
  f_2 (t) = f_1 (t) \int_{t_0}^t \frac{d\tau}{f_1^2 (\tau)
\frac{\partial^2 L_0 (\tau)}{\partial\dot{q}^2}},    \eqno(3.15)
$$
$$
  \xi (t) = f_1 (t) \int_{t_0}^t \frac{d\tau}{f_1^2 (\tau)
\frac{\partial^2 L_0 (\tau)}{\partial\dot{q}^2}} \int_{\tau_0}^\tau
\left[ \frac{\partial L_0 (\eta)}{\partial q} -
\frac{d}{d\eta}\left(\frac{\partial L_0 (\eta)}{d\dot{q}} \right) \right]
f_1 (\eta) d\eta,                                  \eqno(3.16)
$$
where $t_0$ and   $\tau_0$ are already incorporated into constants
$C_1$ and $C_2$ in (3.3)-(3.5).

In the next section we will use also quadratic Lagrangian of the form
$$
   L(\dot{y},y,t) = \frac{1}{2}\frac{\partial^2 L_0}{\partial\dot{q}^2}\dot{y}^2
   +   \frac{\partial^2 L_0}{\partial\dot{q}\partial q}\dot{y}y
   +   \frac{1}{2}\frac{\partial^2 L_0}{\partial q^2} y^2,
                                                     \eqno(3.17)
$$
where coefficients in the expansion are those of (3.1).

The corresponding Euler-Lagrange equation of motion is
$$
 \frac{\partial^2 L_0}{\partial\dot{q}^2} \ddot{y} +
 \frac{d}{dt} \left( \frac{\partial^2 L_0}{\partial\dot{q}^2} \right) \dot{y}
 +  \left[ \frac{d}{dt} \left( \frac{\partial^2 L_0}
{\partial\dot{q}\partial q}
\right) -  \frac{\partial^2 L_0}{\partial q^2}  \right] y = 0
 \eqno(3.18)
$$
with general solution
$$
 y(t) = D_1 f_1(t) + D_2 f_2(t),   \eqno(3.19)
$$
where $f_1(t)$ and $f_2(t)$ remain to be the same previous two solutions.
Denoting
$y'=y(t')$ and    $y''=y(t'')$, constants of integration $D_1$ and  $D_2$
are:
$$
   D_1 \equiv D_1 (t'',t') = \frac{y'f''_2 - y''f'_2}
{f''_2 f'_1 - f''_1 f'_2},     \eqno(3.20)
$$
$$
   D_2 \equiv D_2 (t'',t')  =  \frac{y''f'_1 - y'f''_1}
{f''_2f'_1 - f''_1f'_2}.   \eqno(3.21)
$$

Using the equation of motion (3.18), the Lagrangian (3.17) becomes
$$
  L(\dot{y},y,t) = \frac{1}{2}\frac{d}{dt}
  \left[ \frac{\partial^2L_0}{\partial\dot{y}^2} \dot{y}y +
  \frac{\partial^2L_0}{\partial\dot{y}\partial y} y^2   \right],
                                                              \eqno(3.22)
$$
where $y(t)$ is now related to (3.19).
Since $D_1(t'',t')$ and   $D_2(t'',t')$ are linear in $y''$  and $y'$,
the corresponding classical action
$\bar{S}(y'',t'';y',t') = \int_{t'}^{t''} L(\dot{y},y,t)dt$
is quadratic in $y''$ and
$y'$. An analogous integration to the previous one gives
$$
 \bar{S}(y'',t'';y',t') = \frac{1}{2}\frac{\partial^2\bar{S}_0}
 {\partial y''^2} y''^2 + \frac{\partial^2 \bar{S}_0}{\partial y''
  \partial y'} y'' y' + \frac{1}{2}\frac{\partial^2 \bar{S}_0}{\partial
  y'^2} y'^2                                                  \eqno(3.23)
$$
where:
$$
  \frac{\partial^2\bar{S}_0}{\partial y''^2} =
  \frac{\partial^2 L_0''}{\partial\dot{y}^2} \frac{\dot{f}_2'' f_1'
  - \dot{f}_1'' f_2'}{\Delta (t'',t')} + \frac{\partial^2 L_0''}
  {\partial\dot{y}\partial y},                        \eqno(3.24)
$$
$$
  \frac{\partial^2 \bar{S}_0}{\partial y''\partial y'} =
 \frac{1}{2} \frac{\partial^2 L_0''}{\partial\dot{y}^2} \frac
 {f_2''\dot{f}_1'' - \dot{f}_2''f_1''}{\Delta(t'',t')} +
 \frac{1}{2} \frac{\partial^2 L_0'}{\partial\dot{y}^2} \frac
 {f_2'\dot{f}_1' - \dot{f}_2'f_1'}{\Delta(t'',t')},     \eqno(3.25)
$$
$$
  \frac{\partial^2\bar{S}_0}{\partial y'^2} =
  \frac{\partial^2 L_0'}{\partial\dot{y}^2} \frac{\dot{f}_2' f_1''
  - \dot{f}_1' f_2''}{\Delta (t'',t')} - \frac{\partial^2 L_0'}
  {\partial\dot{y}\partial y}.                         \eqno(3.26)
$$

In virtue of (3.15) one has
$$
 \dot{f}_2 (t) f_1(t) - \dot{f}_1(t)f_2(t) = \left( \frac{\partial^2
 L_0 (t)}{\partial\dot{q}^2} \right)^{-1}.                 \eqno(3.27)
$$
Using (3.27) one can show the following useful formulae:
$$
 \frac{\partial^2}{\partial y''\partial y'}\bar{S}_0(y'',t'';y',t')
 = - \frac{1}{f_2'' f_1' - f_1'' f_2'},               \eqno(3.28)
$$
$$
 \frac{\partial^2}{\partial y^2}\bar{S}_0(y'',t'';y,t) +
 \frac{\partial^2}{\partial y^2}\bar{S}_0(y,t;y',t') =
 \frac{f_2'' f_1' - f_1'' f_2'}{(f_2'' f_1 - f_1'' f_2)
 (f_2 f_1' - f_1 f_2')}
$$
$$
= -\frac{\frac{\partial^2}{\partial y'' \partial y}\bar{S}_0(y'',t'';y,t)
 \frac{\partial^2}{\partial y \partial y'}\bar{S}_0(y,t;y',t')}
 {\frac{\partial^2}{\partial y'' \partial y'}\bar{S}_0(y'',t'';y',t')}.
                                                             \eqno(3.29)
$$

\bigskip
\noindent
{\bf 4. Evaluation of $p$-Adic and Ordinary Path Integrals}
\smallskip

\noindent
Quantum fluctuations lead to deformations of classical trajectory and
any quantum history may be presented as $q(t) = x(t) + y(t)$, where
$y'=y(t')=0$  and  $y''=y(t'')=0$.  The corresponding Taylor expansion
of $S[q]$ around classical path $x(t)$  is
$$
   S[q] = S[x+y] = S[x] + \frac{1}{2!} \delta^2 S[x] = S[x] +
   \frac{1}{2}\int_{t'}^{t''} \left( \dot{y}\frac{\partial}{\partial \dot{q}} +
   y\frac{\partial}{\partial q}   \right)^2 L(\dot{q},q,t)dt ,  \eqno(4.1)
$$
where we used $\delta S[x] = 0$.
For any $v = \infty, 2,3,\cdots$, we can write
$$
   {\cal K}_v (x'',t'';x',t') = \int
   \chi_v \left( -\frac{1}{h} S[x+y] \right)  {\cal D}y,
    \eqno(4.2)
$$
or in the more explicit form,
$$
   {\cal K}_v (x'',t'';x',t') = \chi_v \left( -\frac{1}{h}
   \bar{S}(x'',t'';x',t') \right)
$$
$$
 \times\int_{y'\to 0,t'}^{y''\to 0,t''}
   \chi_v \left(- \frac{1}{2h} \int_{t'}^{t''} \left( \dot{y}
   \frac{\partial}{\partial\dot{q}} +
    y \frac{\partial}{\partial q} \right)^2 L(\dot{q},q,t)dt
   \right) {\cal D}y,
    \eqno(4.3)
$$
where we used $ y''=y'=0 $ and $S[x] = \bar{S}(x'',t'';x',t')$.

From (4.3) it follows that ${\cal K}_v(x'',t'';x',t')$  has the form
$$
    {\cal K}_v(x'',t'';x',t') = N_v(t'',t') \chi_v \left( -\frac{1}{h}
    \bar{S}(x'',t'';x',t')\right) ,   \eqno(4.4)
$$
where $N_v(t'',t')$ does not depend on end points $x''$ and  $x'$.

To calculate $N_v(t'',t')$, let us note that (4.3) can be rewritten as
$$
   {\cal K}_v (x'',t'';x',t') = \chi_v \left( -\frac{1}{h}
   \bar{S}(x'',t'';x',t') \right) K_v (0,t'';0,t'),     \eqno(4.5)
$$
where $ K_v (0,t'';0,t') =  K_v (y'',t'';y',t')
|_{y''=y'=0}$ and
$$
 K_v (y'',t'';y',t')  =
   \int_{y',t'}^{y'',t''}
   \chi_v\left(- \frac{1}{h} \int_{t'}^{t''}
   \left[ \frac{1}{2}\frac{\partial^2 L_0}{\partial\dot{q}^2}\dot{y}^2
   +   \frac{\partial^2 L_0}{\partial\dot{q}\partial q}\dot{y}y
   +   \frac{1}{2}\frac{\partial^2 L_0}{\partial q^2} y^2 \right]
   \right) {\cal D}y.                                    \eqno(4.6)
$$
According to (4.4) and (4.5) one has
$$
  N_v(t'',t') = K_v(y'',t'';y',t')|_{y''=y'=0},             \eqno(4.7)
$$
where
$$
  K_v(y'',t'';y',t') = N_v(t'',t') \chi_v \left( -\frac{1}{h}
   \left[  \frac{1}{2}\frac{\partial^2\bar{S}_0}
 {\partial y''^2} y''^2 + \frac{\partial^2 \bar{S}_0}{\partial y''
  \partial y'} y'' y' + \frac{1}{2}\frac{\partial^2 \bar{S}_0}{\partial
  y'^2} y'^2 \right] \right).                            \eqno(4.8)
$$
Let us employ now (1.3) and (1.4) to find a suitable expression
for $N_v(t'',t')$. The unitary condition (1.4) now reads:
$$
\int_{{\bf Q}_v} \bar{K}_v(y'',t'';y',t') K_v(y,t'';y',t') dy' =
\delta_v(y'' -y). \eqno(4.9)
$$
Substituting $ K_v(y'',t'';y',t')$ from  (4.8) into (4.9),
we obtain
$$
 |N_v(t'',t')|_\infty^2 \int_{Q_v} \chi_v \left( \frac{1}{2h}
 \frac{\partial^2 \bar{S}}{\partial y''^2} (y''^2 -y^2) +  \frac{1}{2h}
 \frac{\partial^2 \bar{S}}{\partial y''\partial y'} (y'' - y)y' \right) dy'
 = \delta_v(y'' -y).                                       \eqno(4.10)
$$
Using the properties of $\delta_v$-functions, in particular:
$\int_{Q_v}\chi_v (ax) dx = \delta_v (a)$ and $\delta_v (ax) = |a|_v^{-1}
\delta_v (x)$, we have
$$
 |N_v(t'',t')|_\infty^2   \chi_v \left(\frac{1}{2h}
 \frac{\partial^2 \bar{S}}{\partial y''^2} (y''^2 -y^2)\right)
\left\vert \frac{1}{h}\frac{\partial^2 \bar{S}}{\partial y''\partial y'}
\right\vert_v^{-1}
\delta_v(y'' - y)
 = \delta_v(y'' -y).                                       \eqno(4.11)
$$
Performing integration in (4.11) over variable $y$, one obtains
$$
  |N_v(t'',t')|_\infty  =
\left\vert \frac{1}{h}\frac{\partial^2 }{\partial y''
\partial y'}\bar{S}_0(y'',t'';y',t')\right\vert_v^{\frac{1}{2}}.                  \eqno(4.12)
$$
We have now
$$
   N_v(t'',t') =  \left\vert\frac{1}{h} \frac{\partial^2}
   {\partial y''\partial y'}\bar{S}_0(y'',t'';y',t') \right\vert_v^{\frac{1}{2}}
   A_v(t'',t') ,                                         \eqno(4.13)
$$
where $\vert A_v(t'',t')\vert_\infty =1$.   To use the condition
(1.3), it has to be rewritten as
$$
 \int_{{\bf Q}_v} K_v(y'',t'';y,t) K_v(y,t;y',t') dy = K_v(y'',t'';y',t').
                                                            \eqno(4.14)
$$
Inserting (4.8) into (4.14), where $N_v (t'',t')$ has the form (4.13),
we get conditions:
$$
  \left\vert \frac{1}{h}\frac{\partial^2}{\partial y''
\partial y}\bar{S}_0(y'',t'';y,t)
  \right\vert_v^{\frac{1}{2}}    \left\vert \frac{1}{h}
\frac{\partial^2}{\partial y\partial y'}
 \bar{S}_0(y,t;y',t')\right\vert_v^{\frac{1}{2}}  \vert 2\alpha \vert_v^{-\frac{1}{2}}
$$
$$
 = \left\vert \frac{1}{h}  \frac{\partial^2}{\partial y'' \partial y'}
  \bar{S}_0(y'',t'';y',t')\right\vert_v^{\frac{1}{2}},          \eqno(4.15)
$$
$$
  \chi_v\left( -\frac{1}{2h}\left[ \frac{\partial^2}{\partial y''^2}
\bar{S}_0(y'',t'';y,t) y''^2  + \frac{\partial^2}{\partial y'^2}
\bar{S}_0(y,t;y',t')     \right]  \right)
$$
$$
  \times\chi_v \left( \frac{1}{2h} \frac{\left( \frac{\partial^2}
{\partial y''\partial y}\bar{S}_0(y'',t'';y,t) y'' +
\frac{\partial^2}{\partial y \partial y'} \bar{S}_0(y,t;y',t') y' \right)^2}
{\frac{\partial^2}{\partial y^2}\bar{S}_0(y'',t'';y,t) +
\frac{\partial^2}{\partial y^2}\bar{S}_0(y,t;y',t') }   \right)
$$
$$
 = \chi_v \left( -\frac{1}{h} \left[ \frac{1}{2}y''^2 \frac{\partial^2}
 {\partial y''^2}  + y'' y'\frac{\partial^2}
{\partial y'' \partial y'}  +
\frac{1}{2} y'^2\frac{\partial^2}{\partial y'^2}
      \right]\bar{S}_0(y'',t'';y',t')  \right),         \eqno(4.16)
$$
$$
   A_v(t'',t) A_v(t,t')\lambda_v (\alpha) = A_v(t'',t') ,    \eqno(4.17)
$$
where
$$
    \alpha = - \frac{1}{2h} \left[ \frac{\partial^2}
{\partial y^2}\bar{S}_0(y'',t'';y,t)
    +\frac{\partial^2}{\partial y^2}\bar{S}_0(y,t;y',t')  \right] . \eqno(4.18)
$$

The condition (4.15) is satisfied due to the property (3.29). The equation
(4.16) follows as the equality between arguments of additive characters,
and such equality of arguments is a consequence of the formulae (3.24)-(3.29).

Solution of the equation (4.17) gives us an explicit expression for
$A_v(t'',t')$. We analyse (4.17) separately for real and $p$-adic cases.
Using definition (2.9) of the $\lambda_\infty$-function in the form
$\lambda_\infty (x) = \exp{\left(-\frac{i\pi}{4} \right)}
\sqrt{\mbox{sign}\ x}$, and analogue of the
first property in (2.7), one can easily show that
$$
 \lambda_\infty \left( \frac{a b}{c}  \right) =
 \lambda_\infty \left( \frac{c}{a b}  \right) =
 \frac{\lambda_\infty (-c)}{\lambda_\infty (-a) \lambda_\infty (-b)}.
                                                   \eqno(4.19)
$$
Using the property (3.29) and then (4.19), we find that
$$
 A_\infty(t'',t') = \lambda_\infty \left( -\frac{1}{2h}
 \frac{\partial^2}{\partial y'' \partial y'}\bar{S}_0(y'',t'';y',t') \right).
                                                   \eqno(4.20)
$$

In order to find $A_p(t'',t')$ let us recall that we consider
quadratic Lagrangian (3.17) with coefficients which are analytic
functions of time $t$, and the corresponding analytic classical solutions
$f_1(t)$  and  $f_2(t)$ of the equation of motion (3.18). Let
these solutions be:
$$
 f_1(t) = a_0 + a_1 t +a_2 t^2 +\cdots,                     \eqno(4.21)
$$
$$
 f_2(t) = b_1 t + b_2 t^2 +\cdots,                         \eqno(4.22)
$$
with the region of convergence which does not exceed the disk
 $|t|_p < |2|_p$.
According to (4.21)  and (4.22) we have:
$$
 f_2''f_1' - f_1''f_2' = a_0b_1 (t'' - t')+ a_0b_2 (t''^2 -t'^2) + \cdots,
                                                             \eqno(4.23)
$$
$$
 f_2''f_1 - f_1''f_2 = a_0b_1 (t'' - t)+ a_0b_2 (t''^2 -t^2) + \cdots,
                                                             \eqno(4.24)
$$
$$
 f_2f_1' - f_1f_2' = a_0b_1 (t - t')+ a_0b_2 (t^2 -t'^2) + \cdots.
                                                             \eqno(4.25)
$$
Note that $\lambda_p(x)$, defined by (2.5) and (2.6),
depend only on the first term in expansion of $x$ if $p\neq 2$, and
the second and third terms if $p=2$. Hence, it is enough to take into account
the first two terms in  expansions (4.23)-(4.25) when their left-hand sides
occur to be arguments of functions $\lambda_p$.
For example, we have
$$
 \lambda_p\left(-\frac{1}{2h} [f_2''f_1 - f_1''f_2 + f_2f_1' -f_1f_2'] \right)
= \lambda_p\left(-\frac{1}{2h} [a_0b_1(t''-t') + a_0b_2(t''^2 -t'^2)] \right)
$$
$$
= \lambda_p \left( -\frac{1}{2h}\frac{\partial^2}{\partial y''\partial y'}
\bar{S}_0(y'',t'';y',t') \right).                             \eqno(4.26)
$$
Using expressions (3.28)-(3.29), properties (2.7) of $\lambda_p$ functions,
and the above expansions of $f_1(t)$ and $f_2(t)$, we get
$$
 A_p(t'',t') = \lambda_p \left( -\frac{1}{2h}
       \frac{\partial^2}{\partial y''\partial y'}
       \bar{S}_0(y'',t'';y',t') \right).            \eqno(4.27)
$$

Since  $\bar{S}_0$, and terms linear in $x''$ and $x'$ of the classical action
(3.7) do not affect function $N_v(t'',t')$,
results obtained for pure quadratic action $\bar{S}_0(y'',t'';y',t')$ (3.23)
are also valid for general quadratic case $\bar{S}_0(x'',t'';x',t')$ (3.7).

In virtue  of the above evaluation one can formulate the following

\noindent
{\bf Theorem}. {\it The $v$-adic kernel ${\cal K}_v(x'',t'';x',t')$ of
the unitary evolution operator, defined by (1.1) and evaluated as the
Feynman path integral,
for quadratic Lagrangians (3.1) (and consequently, for quadratic classical
actions (3.7)) has the form}
$$
{\cal K}_v(x'',t'';x',t') = \lambda_v\left( -\frac{1}{2h}
\frac{\partial^2}{\partial x'' \partial x'}\bar{S}_0(x'',t'';x',t') \right)
\left\vert \frac{1}{h} \frac{\partial^2}
{\partial x'' \partial x'}\bar{S}_0(x'',t'';x',t') \right\vert_v^{\frac{1}{2}}
$$
$$
\times\chi_v\left( -\frac{1}{h} \bar{S}(x'',t'';x',t')  \right)
                                                      \eqno(4.28)
$$
{\it and satisfies the general properties (1.3)-(1.5)}.

\noindent
{\bf Proof}. The formula (4.28) is a result of the above analytic evaluation,
and one has to show that this expression satisfies (1.3)-(1.5). In fact, it is
already shown that (1.3) and (1.4) are satisfied for the reduced
Lagrangian (3.17), and in the analogous way the proof extends to the general
case (3.1). The property (1.5) follows from
$$
 \lim_{t'\to t''} \lambda_v \left( -\frac{1}{2h} \frac{\partial^2 \bar{S}}
 {\partial x'' \partial x'}  \right) \left\vert  \frac{1}{h} \frac{\partial^2
 \bar{S}}{\partial x'' \partial x'}  \right\vert_v^{\frac{1}{2}}   \chi_v \left[
\frac{1}{2h} \frac{\partial^2 \bar{S}}{\partial x'' \partial x'}
\left( x''^2 - 2x''x' + x'^2 \right)  \right] =\delta_v (x'' - x').
 \eqno(4.29)
$$

Starting from (4.28) and using definition (2.9) for $\lambda_\infty$-function
one can rederive well-known result in ordinary quantum mechanics:
$$
{\cal K}_\infty(x'',t'';x',t') = \left( \frac{i}{h}
\frac{\partial^2}{\partial x'' \partial x'}\bar{S}_0(x'',t'';x',t')
\right)^{\frac{1}{2}}
\exp\left( \frac{2\pi i}{h} \bar{S}(x'',t'';x',t')  \right).   \eqno(4.30)
$$

\bigskip
\noindent
{\bf 5. Adelic Path Integrals}
\smallskip

\noindent
In order to introduce adelic path integrals, let us start with analogue
of (1.1) related to eigenfunctions in adelic quantum mechanics,{\it i.e.}
$$
 \psi_{{\bf S},\alpha}(x'',t'') = \int_{\cal A}
 {\cal K}_{\cal A}(x'',t'';x',t') \psi_{{\bf S},\alpha}(x',t') dx',  \eqno(5.1)
$$
where  $\psi_{{\bf S},\alpha}(x,t)$ has the form (2.16), and  adelic
propagator ${\cal K}_{\cal A}(x'',t'';x',t')$  does not depend on ${\bf S}$.
Since the equation (5.1) must be valid for any set ${\bf S}$ of primes $p$,
and adelic eigenstate is an infinite product of real and $p$-adic
eigenfunctions, it is natural to consider adelic propagator in the following
form:
$$
 {\cal K}_{\cal A}(x'',t'';x',t') =
 {\cal K}_\infty(x_\infty'',t_\infty'';x_\infty',t_\infty')  \prod_p
 {\cal K}_p(x_p'',t_p'';x_p',t_p'),                                \eqno(5.2)
$$
where $ {\cal K}_\infty(x_\infty'',t_\infty'';x_\infty',t_\infty')$  and
$ {\cal K}_p(x_p'',t_p'';x_p',t_p')$ are propagators in ordinary and
$p$-adic quantum mechanics, respectively.

From (5.2) we see that one can introduce adelic path integral as an
infinite product of ordinary and $p$-adic path integrals for all primes
$p$. Intuitively, we  regard adelic Feynman's functional integrals
as path integrals on adelic spaces. Now we can consider
(5.2) in the form of path integrals, and let us
symbolically write the left-hand side as
$$
  {\cal K}_{\cal A}(x'',t'';x',t') = \int_{x',t'}^{x'',t''}
  \chi_{\cal A} \left( -\frac{1}{h} S_{\cal A}[q] \right)
  {\cal D}_{\cal A}q,                                        \eqno(5.3)
$$
where $\chi_{\cal A}(x)$ is adelic additive character (2.12),
$S_{\cal A}[q]$ and  $ {\cal D}_{\cal A}q $  are adelic action and the
Haar measure, respectively. For practical considerations, we
define adelic path integral in the  form
$$
  {\cal K}_{\cal A}(x'',t'';x',t') = \prod_v
\int_{x_v',t_v'}^{x_v'',t_v''} \chi_v \left( -\frac{1}{h}
\int_{t_v'}^{t_v''} L (\dot{q}_v,q_v,t_v) dt_v  \right)
{\cal D}q_v,                                                  \eqno(5.4)
$$
where index $v=\infty, 2, 3,\cdots, p, \cdots$ denotes real and all
$p$-adic cases. As an adelic Lagrangian one understands infinite
sequence
$$
 L_{\cal A}(\dot{q},q,t) = (L(\dot{q}_\infty,q_\infty,t_\infty),
L(\dot{q}_2,q_2,t_2), L(\dot{q}_3,q_3,t_3),\cdots,
 L(\dot{q}_p,q_p,t_p),\cdots),                               \eqno(5.5)
$$
where $|L(\dot{q}_p,q_p,t_p)|_p \leq 1$ for all primes
$p$ but a finite set $\bf S$ of them. Consequently, an adelic quadratic
Lagrangian looks like (5.5), where each element $L(\dot{q}_v,q_v,t_v)$
has the same form (3.1). Note that to escape an abundance of indices, we
often omit some of them in the cases when they are
implicitly understood.

Taking into account results obtained in the previous section, we can
write adelic path integral for quadratic Lagrangians (and consequently,
quadratic classical actions) as
$$
{\cal K}_{\cal A}(x'',t'';x',t') = \prod_v
\lambda_v\left( -\frac{1}{2h}
\frac{\partial^2}{\partial x_v'' \partial x_v'}\bar{S}_0(x_v'',t_v'';x_v',t_v')
\right)   \left\vert \frac{1}{h} \frac{\partial^2}
{\partial x_v'' \partial x_v'}\bar{S}_0(x_v'',t_v'';x_v',t_v')
\right\vert_v^{\frac{1}{2}}
$$
$$
\times\chi_v\left( -\frac{1}{h} \bar{S}(x_v'',t_v'';x_v',t_v')  \right).
                                                      \eqno(5.6)
$$

Note that vacuum state $\Omega(|x_p|_p)$ transforms as
$$
 \Omega(|x_p''|_p) = \int_{{\bf Q}_p}{\cal K}_p(x_p'',t_p'';x_p',t_p')
 \Omega(|x_p'|_p) dx_p' =
 \int_{{\bf Z}_p}{\cal K}_p(x_p'',t_p'';x_p',t_p') dx_p'.     \eqno(5.7)
$$
As a consequence of (5.7) one has
$$
 \int_{{\bf Z}_p}{\cal K}_p(x_p'',t_p'';x_p,t_p) {\cal K}_p(x_p,t_p;x_p',t_p')
dx_p  = {\cal K}_p(x_p'',t_p'';x_p',t_p'),     \eqno(5.8)
$$
which may be regarded as an additional condition on $p$-adic path integrals
in adelic quantum mechanics for all but a finite number of primes $p$.
Conditions (5.7) and (5.8) impose a restriction on  a dynamical system
to be adelic (see, \cite{brank4}). It is practically a restriction on
time $t_p$ to have consistent adelic time $t$.

\bigskip

\noindent
{\bf 6. Concluding Remarks}
\smallskip

\noindent
Evaluating the method of Feynman's functional integral simultaneously on
real and $p$-adic one-dimensinal spaces, in the previous sections we
derived general expressions for propagators ${\cal K}(x'',t'';x',t')$
in ordinary, $p$-adic and adelic quantum mechanics. Especially, it
has been done for Lagrangians $L(\dot{q},q,t)$ which are polynomials
at most the second degree in  dynamical variables $\dot{q}$ and $q$.

It is worth pointing out that the formalism of ordinary and $p$-adic path
integrals can be regarded as the same at different levels of evaluation,
and the obtained results have the same form. In fact, this property
of number field invariance has to be natural for general mathematical
methods in physics and fundamental physical laws ({\it cf.}
\cite{volov4}).
In the present case it is mainly a consequence of the form
invariance of the Gauss
integral under interchange of ${\bf R}$ and ${\bf Q}_p$ (see, (2.4)
and (2.8)).

The idea to derive expression (4.27) in the above way was proposed
by one of the authors (B.D.) in contribution to the
Bogolyubov conference \cite{brank5}. Some aspects of $p$-adic path integral
for quadratic actions are considered earlier in Ref. \cite{brank3}.

The above results, obtained for one-dimensinal systems, are an appropriate
starting point for a generalization to two-dimensional \cite{brank6}
and higher-dimensional cases. These results can be also exploited to
construct and investigate a new approach to the Brownian motion
with $p$-adic effects.

\bigskip

{\bf Acknowledgements}. B.D. would like to thank I.V. Volovich for
useful discussions. The work on this paper was partially supported
by the Serbian Ministry of Science, Technologies and Development
under contract No 1426. G.S.Dj. was partially supported by DFG
Project "Noncommutative space-time structure - Cooperation with
Balkan Countries". The work of B.D. was supported in part by RFFI
grant 02-01-01084.

\end{document}